\documentstyle[titlepage]{article}
\newcommand{\qt}{quantum theory}
\newcommand{\qo}{quantum object}  

\begin{document} \title{Teaching the EPR--Paradox at High School ?}
 \author{Gesche Pospiech\\ Institut for Didaktik der
 Physik\\ Johann Wolfgang Goethe University Frankfurt}
\maketitle
\begin{abstract}
 The discovery of quantum mechanics in the beginning of our century
 led to a revolution of physical world view. Modern experiments on the
 border of the classical and the quantum regime made possible by new
 techniques open better insight and understanding of the quantum world
 and have impact on new technological development. Therefore it seems
 important that students and even pupils at higher grades become
 acquainted with the principles of quantum mechanics. A suitable way
 seems to be given by treatment of the EPR--gedankenexperiment.
\end{abstract}
\subsection{Introduction}
The first question to be answered is: why should \qt\ be taught at
school at all? For choosing this topic there are the following three
reasons:
\begin{enumerate}
\item Quantum theory is the fundamental theory of modern physics. It
plays a significant role in nearly all modern developments of
physics. Many recent experiments and research in nanostructures with
large applicability in technology rely on quantum effects.
\item Quantum theory has important philosophical aspects. Many people
are highly interested in interpretation and understanding \qt\ as
shows up in the many popular books about this subject.
\item Pupils at the age from 16+ on are searching for their place in
the world. They are trying to understand the world and are open for
philosophical hints that help them in building their own world view.
\end{enumerate}
But often this highly fascinating subject is avoided at school because
of mathematical and conceptual difficulties. I therefore want to show
a possible way to introduce \qt\ in a manner suitable for interested
pupils. In this article I concentrate on the mathematical part because
here lie some difficulties. For an introduction into the philosophical
aspects I have developed a dialogue between philosophers from
different times - classical antiquity (Parmenides), the Enlightenment
(Kant) and from our century, published elsewhere, (\cite{dpg98}).
The goal of the following is to clarify the main difficulties in
teaching the physical basis of \qt\ and how to keep them minor.  \\
Speaking qualitatively about \qt\ in an adequate way is nearly
impossible in itself since all our concepts and terms have been
developed along everyday experience. Hence our language is well suited
to communicate about concrete physical objects with well determined
properties or about psychological issues. In my opinion this last
property should be used in dealing with the interfering and
superposing objects of \qt\ that may have more similarities or
associations with pyschological feelings than with concrete balls or
waves occurring in classical physics. In every attempt to talk about
\qt\ one has to be aware of this principal difficulty already
recognized by Bohr, Heisenberg, Pauli and others. In the complementary
worlds of the quantum regime and the classical regime we only are at
home in the classical regime. The other regime remains accessible only
through sophisticated experiments - even if an experimentalist would
call them easy and simple.... One way out might be to talk in images -
but soon one arrives at poor analogies. Hence, in order to reach more
than only a superficial knowledge at least some hints to the
mathematical background of \qt\ must be given. On the other hand at
least in the beginning some themes should be avoided to facilitate the
pupils the understanding of the peculiarities of \qt.

\subsection{What can be done without too many technicalities?}
It is nearly impossible to understand \qt\ without considering its
mathematical structure. Nevertheless at school the mathematical
apparatus of quantum mechanics has to be abandoned for the major part.
The main ideas, however, can be presented quite easily with help of
the typical quantum phenomenon ``spin'' having no classical analogues.
Experiments with polarized photons may help in conveying the
essentials. In the following I describe the reasons for taking spin
as the first subject in treating \qt\ in more detail. Furthermore I
explain with the example of EPR--gedanken experiment how to proceed.\\
Treating the phenomenon ``spin'' right in the beginning has several
advantages:
\begin{itemize}
\item Spin lies at the heart of quantum theory. Its properties are
used to explain the different statistics, the fine structure of
spectra, the splitting of spectra in external (electrical or magnetic)
fields. Already a spin system consisting of two particles, i.e. living
in a four--dimensional Hilbert space can no longer be described
classically. This proof is similar to the proof of Bell inequality,
(\cite{baumgaertel}).
\item The procedure to describe spin mainly by its structure is
typical of \qt. Furthermore, the mathematics of spin is quite simple,
using mainly the well--known Pauli-- matrices:
$$\sigma_x=\left(\begin{array}{cc}0& 1\\
1&0\end{array}\right),\qquad\sigma_y=\left(\begin{array}{cc}0& i\\
-i&0\end{array}\right),\qquad\sigma_z=\left(\begin{array}{cc}1& 0\\
0&-1\end{array}\right). $$ With the help of these quite simple looking
matrices, acting on two--dimensional Hilbert space, the most
essential mathematical structures of quantum theory can be explained
and interpreted, see table 1. Some details are explained in the next
section.
\\
\item The meaning of Heisenberg uncertainty relation can be explained
as principal non--existence of fixed values for properties hence of
their non--determination, \cite{diphoton}. Therefore the danger that
the uncertainty is perceived as measurement mistakes can be
drastically diminished. Some helpful constructions even can be
visualized on the blackboard.
\item Spin is a phenomenon of special importance in modern experiments
reaching from Nuclear Magnetic Resonance used in medical applications
to realizations of the Einstein--Podolsky--Rosen gedankenexperiment.
Its treatment opens the way to a discussion of philosophical aspects
of \qt\ which quickly reaches the main points: the question of reality
and objectivity in nature treated on a mathematical and physical
foundation.
\end{itemize}   
\subsubsection{An Example}
As an example I show how the arguments of Einstein, Podolsky and Rosen
in their famous paper \cite{orgepr} can be used to show the power of
the mathematical formalism and - even more important - how the
mathematical constructions can be interpreted in this framework. This
allows a bridge to be built from the mathematical structures over the
physical phenomena and connecting to a philosophical discussion.
Instead of arguing very sophisticated within the mathematical
formalism the main goal should be to uncover the main aspects of \qt\
and in this way to build a solid fundament from which the mathematics
can be developed further, (see also table 1).  \\ The argumens of EPR
can be developed the following way:
\subparagraph{Step 1: The mathematical tools} In 1935, the year of the
EPR-paper the mathematical
framework has just been settled implying the following
main points:\begin{itemize}\item The {\sl state} of a quantum object
is given by a state vector $\psi$ containing all the available
information, i.e.  a complete description of the physical properties
of the quantum object.
\item Each {\sl physical quantity} is given together with all the
possible results of a measurement of that quantity and corresponding
eigenstates, i.e. all the states a quantum object can attain after a
measurement. A mathematical realization of this concept is given for
instance by matrices. \item Arbitrary states can be expressed with aid
of the eigenstates of such a matrix resp. physical quantity.
\end{itemize}
\begin{table}
\begin{tabular}{|p{2.5cm}|p{5cm}|p{4.5cm}|} \hline {\bf Mathematical term}
& {\bf Physical Interpretation} &\bf Example \\ \hline vector&physical
state& $\psi =\left(\begin{array}{c}1\\ 0\end{array}\right)$ \\ \hline
operator&physical quantity&$\sigma_x=\left(\begin{array}{cc}0& 1\\
1&0\end{array}\right)$ \\ \hline eigenvalues of an operator& possible
results of measurements& $+1,-1$ \\ \hline eigenstates of an operator (normalized to 1)&
physical states with a fixed value for the physical quantity in
question&\parbox[b]{4.5cm} { $\left(\begin{array}{c}\frac{1}{\sqrt 2}\\
\frac{1}{\sqrt 2}\end{array}\right)$,
$\left(\begin{array}{c}\frac{1}{\sqrt 2}\\ -\frac{1}{\sqrt
2}\end{array}\right)$ }\\ \hline vector addition& superposition (no
fixed value for the physical quantity in question)& \\ \hline
development into eigenstates&representation of arbitrary physical
states with respect to the corresponding physical quantity&
$\left(\begin{array}{c}1\\ 0\end{array}\right)=\frac{1}{\sqrt 2}\left[
\left(\begin{array}{c}\frac{1}{\sqrt 2}\\ \frac{1}{\sqrt
2}\end{array}\right) +\left(\begin{array}{c}\frac{1}{\sqrt 2}\\
-\frac{1}{\sqrt 2}\end{array}\right)\right]$\\ \hline coefficients
(squared) of development & probability of getting the corresponding
measurement result& probability of getting either $+1$ or $-1$:
$\left(\frac{1}{\sqrt 2}\right)**2=\frac{1}{2}$\\ \hline
\end{tabular}  \end{table}
 In a deviation from the original argument of EPR I would advise
taking the spin realized with the above--mentioned Pauli--matrices as
a concrete example. The students can compute the eigenvalus and
eigenstates easily from the matrices. The possible measurement results
(eigenvalues) are $+ 1$ and $-1$ together with the corresponding
eigenstates. The first expriment showing this property directly has
been the Stern--Gerlach--experiment. Furthermore the Pauli--matrices
fulfill the condition crucial for the next step of argument of EPR:
they do not have any eigenstates in common. Hence there always are
several possibilities to represent the spin state of a quantum object,
namely with respect to the respective eigenstates of the different
matrices correponding to the spin di
rections. The representation of an
arbitrary spin state $\psi(s)$ with respect to the eigenstates of
$\sigma_x$ would be
$$\psi(s)=c_1\left(\begin{array}{c}\frac{1}{\sqrt 2} \\ \frac{1}{\sqrt
2} \end{array}\right)+c_2\left(\begin{array}{c}\frac{1}{\sqrt 2} \\
-\frac{1}{\sqrt 2} \end{array} \right)$$ and with respect to the
eigenstates of $\sigma_z$ the same state $\psi$ would look like:
$$\psi(s)=k_1\left(\begin{array}{c}1\\
0\end{array}\right)+k_2\left(\begin{array}{c}0\\ 1\end{array}
\right)$$ with different coefficients $(c_i)\not=(k_i)$. (For a
concrete example look at the table.) This fact perhaps does not matter
too much since we always can change coordinates (here it would be the
rotation of a coordinate system by an angle of 45 degree). But here
it means that the spin state of a given system possesses two different
representations belonging to the ``same piece of reality''(EPR). The
most interesting thing happens in the next step!
\subparagraph{Step 2: The experimental setup} 
Two \qo s, e.g. photons, are brought into interaction or produced in a
single process and hence become entangled i.e. they share a common
``history''. After that they are separated from each other without any
further manipulation, let us say one is brought to the moon, the
second stays on earth. \\ Because of their common ``history'' they are
described by one common state $\psi$ which is not just the addition of
the states of the single photons. This consideration is central for
the whole argument of EPR. The development of the entangled state of
both photons into eigenstates with respect to eigenstates of
$\sigma_x$ is given by:
$$\psi(s_1,s_2)=\psi_1(s_1)\left(\begin{array}{c}1\\
1\end{array}\right)+\psi_2(s_1)\left(\begin{array}{c}1\\ -1\end{array}
\right)$$ and with respect to the
eigenstates of $\sigma_z$:
$$\psi(s_1,s_2)=\phi_1(s_1)\left(\begin{array}{c}1\\
0\end{array}\right)+\phi_2(s_1)\left(\begin{array}{c}0\\ 1\end{array}
\right)$$The only difference to the representations above is that the
coefficients now depend on $s_1$ The meaning of these two
representations is that {\sl photon 1 is described differently
depending on the description chosen for photon 2}, namely
$\psi_i(s_1)$ resp. $\phi_i(s_1)$. This is called the entanglement of
the two photons. Therefore I would prefer to call the whole system
consisting out of these two photons rather a ``diphoton'' in order to
emphasize that they build {\sl one whole} (also see step 5 below).
\subparagraph{Step 3: Classical Assumptions} Assuming a fixed objective
reality and demanding that physics has to give a complete description
of reality Einstein arrives at a contradiction to the predictions of
\qt. More precisely, Einstein assumes:
\begin{enumerate}
\item Separability\\ Classical Physics only knows action between
objects in direct contact with each other. With ``object'' in this
sense I also denote e.g. fields. Hence if two objects are separated in
space, including intermediating fields, all future manipulations on
them are absolutely independent from each other. We could summarize
this in the sentence: Spatially separated objects also are physically
separated. This is an implicit assumption of EPR that is not spoken
out directly, but is underlying the whole argument as can be seen in
the last paragraph of the famous EPR-paper \cite{orgepr}. \\
``Separability'' hence means that the respective descriptions of two
spatially separated photons should be totally independent from each
other.
\item Physical Reality\\ Einstein defines a pragmatic criterion for
reality: Every well determined physical quantity has to have a
representation in the theory. The point herein lies in the question:
Which properties are well determined? Einstein regarded every physical
quantity that can be measured as well determined. But \qt\ deviates in
so far from classical physics as not all (in principle) measurable
quantites have well determined properties at the same instant. They
only possess them as a {\sl potentiality}.
\end{enumerate}
From this view point the different descriptions from above (step 2)
should not occur in a ``good'' physical theory.
\subparagraph{Step 4: Quantum Theoretical Outcome} We can get information
about the photons only after a measurement. What can possibly happen
then? There are several possibilities (as an example):
\begin{enumerate}\item The spin of photon 2 is measured in
$x$-direction. At the same instant the spin state of photon 1 is
$\psi_1(s_1)$ or $\psi_2(s_1)$ according to the result of the
measurement at photon 2. \item The spin of photon 2 is measured in
$z$-direction. At the same instant the spin state of photon 1 is
$\phi_1(s_1)$ or $\phi_2(s_1)$ according to the result of the
measurement performed on photon 2.\end{enumerate}That means that
photon 1 immediately ``knows'' the {\bf kind} of measurement done on
photon 2 far away as well as its {\bf result}. Einstein calls this a
``spooky action at a distance'', which may not occur in classical
physics.
\subparagraph{Step 5: Interpretation}
The behaviour of both entangled photons is strongly connected to each
other, they behave in spite of their spatial separation as one single
quantum object. Therefore I propose to call these both a ``diphoton''
which suggests more clearly that there is only {\sl one common state}
of the whole system, and not an addition of states of separated
photons. Furthermore, the outcome of measurements demonstrates that we
may not assume that photon 1 or photon 2 had fixed values for their
spin directions before measurement. For this purpose one could use the
comfortable Dirac--notation for spin states e.g.:
$\psi(s_1,s_2)=|1,0\rangle-|0,1\rangle$ for an entangled spin state
instead of the above used vector--notation. The Dirac--notation has
the advantage of showing only the {\bf relative} directions of spins
of both photons, which is the only property that is fixed and well
determined (in absence of manipulations). The directions themselves
are not determined, they only show up {\bf after} a measurement. Fixed
values of properties do not exist in general, they only emerge in
measurements. Once this essential point is grasped the way is open for
applications. \\ \\ This access consequently avoids possible pitfalls
which in general erschweren understanding \qt.
\subsection{Which Themes to Avoid in a First Approach?} 
From historical reasons, having their roots in the development of \qt,
most ways of teaching the concepts of quantum physics refer to
classical models. This ``procedere'' causes principal difficulties in
understanding. Therefore every reference to classical concepts should
be avoided as far as possible. The most important points to avoid are:
\begin{itemize}
\item {\bf Speaking about position and momentum, i.e. about
trajectories} \\ If the concepts of position and momentum ---
well--known from everyday experience --- are used at the very
beginning of a course in \qt\ there exists the danger of transferring
classical thinking to \qt, although everybody would say: clearly, in
\qt\ there are no trajectories. Conceptual difficulties arising from
use of the terms `` position'' and ``velocity'' can be avoided in the
simplest manner if these fundamental classical concepts do not play
any role in the beginning of a course in \qt. Then any association of
classical ideas might disappear and students might recognize the
philosophical significance of \qt\ far more easily. The most prominent
example is the famous Heisenberg uncertainty relation for position and
momentum which easily is misunderstood in a sense that the uncertainty
simply relies on disturbance by measurement in the usual sense.
Instead the uncertainty relations are kind of measure for
distinguishing classical behaviour from quantum behaviour in that they
determine whether two physical quantities can attain fixed values at
the same instant. If two physical quantities can attain fixed values
at the same instant the quantum object in question behaves
``classically'', if not it displays quantum behaviour as e.g. spin.
Hence the role and the implications of the non--existence of fixed
values for some properties at the same instant - as expressed in
uncertainty relations - might not be fully appreciated in their
revolutionary potential if one concentrates on ``position'' and
``momentum''.  Besides undesired analogies to Newtonian mechanics the
corresponding operators for position and momentum and their
eigenstates are mathematically far more difficult to handle than the
$2\times2$--spin--matrices.
\item {\bf Speaking about particle--wave--dualism}\\ Waves and
particles both are classical concepts, complementary to each other.
One could illustrate their relation by looking at the same object from
different sights. E.g. a cylinder standing upright appears completely
different from the above (a circle) compared to a look from the side
(a rectangle). But this observation does not meet the essential point
in \qt. \\ A first step to avoid analogy to classical phenomena would
be to use the term ``\qo'' instead of wave or particle. Only after the
quantum mechanical concepts are fixed there might be a careful use of
those ``classical'' terms be allowed where unevitable. Perhaps the
importance of using suitable terms may become clear with the example
of the double--slit--experiment. If it is replaced by the so called
Taylor--experiment in which photons display {\underline {at the same
time}} wave properties - they show interference - as well as particle
properties - they arrive at distinct points on the film, the necessity
of changing concepts gets far more obvious.
\item {\bf Speaking about spin as sort of spinning around}\\ One
should not give an image of spin. Especially one may not think in
terms of an electron spinning around. The quantum mechanical spin is
simply structure manifesting itself and its behaviour through
experiments, especially in the Stern--Gerlach--experiment which may
serve as an introductory experiment. As shown above the abstract
structure of spin can de introduced to a certain extent, depending on
the mathematical capabilities of students.
\end{itemize}
Those three points are mentioned here because their avoidance
breaks with the tradition of teaching and speaking about \qt. The
preceding sections showed an alternative.
\subsection{Conclusion and Perspectives}
The recent EPR--experiments are the starting point for all the current
developments concerning the fundamentals of \qt\ as well as
technological utopies in the area of quantum computing and
teleportation. In addition it widely opens the door to philosophical
discussions. In so far the EPR--gedanken experiment lies at the heart
of \qt\ and its interpretation.  \\ The entrance to quantum mechanics
with help of the phenomenon spin quickly gives gifted or interested
pupils a possibility of discussing the properties of \qo s, the
mathematical structures and the interpretation of quantum mechanics
on a technically very modest level but nonetheless quite precise.
As the spin inevitably is a purely quantum mechanical phenomenon this
opens via the EPR--gedanken--experiment a short way into the crucial
points of understanding concepts as well as philosophical implications
of \qt\ and hence gives the possibility for people to revisit their
view of nature, their {\sl Weltbild}. I regard this an important
contribution to general education. \bibliographystyle{alpha}
\bibliography{lit}

\begin{thebibliography}{EPR35}

\bibitem[Bau]{baumgaertel}
Hellmut Baumg"artel.
\newblock {M}athematische {G}rundlagen der {Q}uantentheorie.
\newblock manuscript, http://www.math.uni-potsdam.de/mp1/vorlesun.htm.

\bibitem[EPR35]{orgepr}
Albert Einstein, Boris Podolsky, and Nathan Rosen.
\newblock Can quantum-mechanical description of physical reality be considered
  complete?
\newblock {\em Physical Review}, 48:696--702, 1935.

\bibitem[Pos98]{dpg98}
Gesche Pospiech.
\newblock Vom {A}tommodell zur {Q}uantenphysik und zur"uck.
\newblock In {F}achverband {D}idaktik der~{P}hysik {D}eutsche~{P}hysikalische
  {G}esellschaft, editor, {\em {D}idaktik der {P}hysik -- {B}eiträge zur 62.
  {P}hysikertagung {R}egensburg 1998}, pages 178--183, 1998.

\bibitem[Pos99]{diphoton}
Gesche Pospiech.
\newblock Spukhafte {F}ernwirkungen in der {Q}uantentheorie?
\newblock {\em {P}hysik in der {S}chule}, 37:56--59, 1999.

\end{thebibliography}
\end{document}